\documentclass[aps,prd,floatfix,twocolumn,superscriptaddress,showpacs,preprintnumbers,nofootinbib,10pt]{revtex4}
\usepackage{epsfig}
\usepackage{dcolumn}
\usepackage{bm}
\usepackage{amsmath}
\usepackage{amsfonts}
\usepackage{amssymb}
\usepackage{subfigure}
\usepackage{graphicx}
\usepackage{latexsym}
 \usepackage{slashed}
 \usepackage{verbatim}
 \usepackage{ulem}
\usepackage{color}
\usepackage{amsmath}
\usepackage{amsfonts}
\usepackage{amssymb}
\usepackage{graphicx}
\usepackage{caption}
\usepackage{subfigure}
\usepackage{slashed}
\usepackage{hyperref}
\def\nnd{\end{document}}

\def\be{\begin{equation}}
\def\ee{\end{equation}}

\newcommand{\bea}{\begin{eqnarray}}
\newcommand{\eea}{\end{eqnarray}}
\newcommand{\bwt}{\begin{widetext}}
\newcommand{\ewt}{\end{widetext}}

\def\u
\def\hZ{\widehat Z}

\def\eed{\end{document}}

\def\m_z{m_{\textrm {Z}}}

\renewcommand{\u}{\rm{u}}

\def\be{\beta}

\def\rm#1{\textrm{#1}}

\makeatother
\begin{document}

\title{Bottom quark contribution to spin-dependent dark matter detection}
\preprint{ADP-15-20-T922}

\author{Jinmian Li}
\email[E-mail: ]{jinmian.li@adelaide.edu.au}
\affiliation{ARC Centre of Excellence for Particle Physics at 
the Terascale (CoEPP),\\
and CSSM, Department of Physics, University of Adelaide, 
South Australia 5005, Australia}

\author{Anthony W. Thomas}
%\email[E-mail: ]{Anthony.Thomas@adelaide.edu.au}
\affiliation{ARC Centre of Excellence for Particle Physics at 
the Terascale (CoEPP),\\
and CSSM, Department of Physics, University of Adelaide, 
South Australia 5005, Australia}

\begin{abstract} 
We investigate a previously overlooked bottom quark contribution to 
the spin-dependent cross section for Dark Matter(DM) scattering from 
the nucleon. While the mechanism is 
relevant to any supersymmetric extension of the Standard Model, for 
illustrative purposes we explore the consequences within 
the framework of the Minimal Supersymmetric Standard Model(MSSM).  
We study two cases, namely those where the DM is predominantly 
Gaugino or Higgsino.  In both cases, there is a substantial, viable 
region in parameter space 
($m_{\tilde{b}} - m_\chi \lesssim \mathcal{O}(100)$ GeV) 
in which the bottom contribution becomes important. We  show that a 
relatively large contribution from the bottom quark is consistent 
with constraints from spin-independent DM searches, as well as some incidental model dependent constraints.
\end{abstract}


\maketitle

\section{Introduction}
\label{sec:intro}
The existence of non-baryonic Dark Matter(DM) has been established 
by many astronomical observations~\cite{Bertone:2004pz,Komatsu:2010fb}. 
Amongst the many candidates for DM, the so-called Weakly Interacting 
Massive Particles(WIMPs), which would have a mass  in the range $\mathcal{O}(1)$ GeV $-\mathcal{O}(1)$ TeV, are one of the most attractive. 
These particles would only interact with Standard Model(SM) 
particles through weak interactions (and gravity), in order  
to yield a DM relic density consistent with  
measurement $\Omega_{\text{DM}} h^2 = 0.1199\pm 0.0027$~\cite{Ade:2013zuv}. 

Direct detection of DM relies on observing the recoil energy after 
scattering from normal matter through weak interactions. 
Several DM direct detection experiments have claimed a 
possible excess, namely DAMA~\cite{Bernabei:2010mq}, 
CoGeNT~\cite{Aalseth:2010vx},CRESST~\cite{Angloher:2011uu} and 
CDMS~\cite{Agnese:2013rvf}. On the other hand, these results 
are challenged by the absence of signals at XENON100~\cite{Aprile:2012nq} 
and LUX~\cite{Akerib:2015rjg}, as well as CDMSlite~\cite{Agnese:2015nto} in the light DM region. The coherent, spin-independent(SI) interaction 
between a DM particle, generically labelled $\chi$, and a nucleus 
is proportional to the nucleon number.   Because of the relatively 
heavy nuclei chosen for most of the above mentioned experiments, both the 
observed excess and stringent exclusion limits are based 
on SI $\chi-p$ scattering.  

As for spin-dependent(SD) 
DM detection~\cite{Ellis:1987sh}, in a simple shell model 
the spin of the nucleus is that of a single, unpaired nucleon.
As a consequence, the matrix element for SD $\chi$-nucleus scattering 
will be roughly comparable with that for SI $\chi$-nucleon scattering, 
with no enhancement by the nucleon number.  As a result, 
the current DM direct searches place only very loose bounds on 
the SD cross section~\cite{Aprile:2013doa,Felizardo:2011uw,Aartsen:2012kia}. 

In the standard calculation of SD DM-nucleon 
scattering the heavy quark contribution 
is usually neglected. That is, only the 
contributions from $\Delta u, \, \Delta d$ and 
$\Delta s$ are included. However, as explained in the 
context of the proton weak charge~\cite{Bass:2002mv}, 
the usual decoupling of heavy quarks through the 
Appelquist-Carrazone theorem~\cite{Appelquist:1974tg} 
does not apply to quantities influenced by the U(1) axial 
anomaly~\cite{Adler:1969gk,Crewther:1974vw,Collins:1978wz,Kaplan:1988ku,Carlitz:1988ab}. 
In that case, 
rather than being suppressed by inverse powers of the heavy 
quark mass, the suppression
is only logarithmic. These logarithmic corrections were studied in considerable 
detail by Bass {\it et al.} in 
Refs.~\cite{Bass:2002mv,Bass:2005ku,Crewther:2005th}, 
at both leading and next-to-leading order. 
As we shall explain here, there are 
interesting scenarios of supersymmetry(SUSY), generally involving a relatively light 
sbottom, where the logarithmic radiative correction involving 
the $b$-quark that is further enhanced by resonant effect may make a significant contribution to SD DM-nucleon scattering.

%{\color{blue}Once it is realized that the triangle diagram involving a $b$-quark is only 
%logarithmically suppressed, it is apparent that this contribution may 
%be enhanced in many supersymmetric models.}
Indeed, SUSY~\cite{Nilles19841,Haber198575} is widely believed  
to provide the most 
promising explanation for new physics beyond SM. In SUSY models with R-parity 
conservation, the lightest supersymmetric particle(LSP) is stable and 
can become a DM candidate. On the other hand, both the LHC SUSY 
searches~\cite{atlas8,cms8} and naturalness 
arguments~\cite{Barbieri:1987fn,Ellis:1986yg} suggest that only the third 
generation supersymmetric quarks(squarks) can be light. 
In Ref.~\cite{Batell:2013psa}, it has been argued that an 
sbottom with a mass as light as $\sim \mathcal{O}(15)$ GeV might still 
be consistent with current searches. 
In other models, such as the simplified model framework~\cite{Abdallah:2014hon} 
and flavored DM models~\cite{Batell:2011tc,Agrawal:2011ze}, 
the DM can only couple 
to the bottom quark, as motivated by the recent DM indirect 
signals~\cite{Agrawal:2014una}. 
Studying the bottom quark contribution to the DM-nucleon 
SD cross section is crucial 
in models of this type.

In this work, we focus on the Minimal Supersymmetric Standard Model (MSSM) 
with a relatively light sbottom, showing when and how the bottom contribution 
becomes important.  When the DM is Wino, there is no coupling between DM and 
the $Z$-boson and only squark mediated processes can contribute 
to $\chi$-nucleon scattering. We investigate the parameter space where the 
sbottom  contribution is comparable to, or larger than, the first generation 
squark contribution.   
When the DM is Higgsino, the first two generation squark mediated processes 
are greatly suppressed by their small Yukawa couplings. 
However, the Higgsino can 
couple to the $Z$-boson. The constructive and destructive interference effects 
between $Z$ and sbottom ($\tilde{b}$) mediated processes are discussed in 
detail for a number of variations on the structure of the neutralino.

Any sbottom mediated process that contributes to the SD scattering cross section 
can also contribute to SI scattering. We consider the stringent LUX constraint 
on SI DM detection for light sbottom scenarios of interest. 
A relatively large SD bottom contribution can indeed be found, while 
maintaining consistency with the LUX constraint.
We also consider several model dependent constraints from collider searches.
We stress that our conclusion has implications beyond the MSSM, which 
is used here purely for purposes of illustration.

This paper is organised as follows. In Sec.~\ref{sec:eft}, we explain 
the theoretical framework for the calculation of the SD DM-nucleon scattering 
cross section. Sec.~\ref{sec:sbsd} discusses the bottom contribution for 
Wino and Higgsino DM. The corresponding SI detection and LHC constraints on the light sbottom scenario are considered in Sec.~\ref{sec:si} and Sec.~\ref{sec:cons}.
In Sec.~\ref{sec:conc} we present some concluding remarks. 

\section{Effective interaction for Spin-Dependent DM-nucleon scattering in MSSM}
\label{sec:eft}
Given a general effective Lagrangian
\begin{align}
\mathcal{L}^{\text{eff}}_{\text{SD}} &= d_q \bar{\chi} 
\gamma^\mu \gamma_5 \chi \bar{q} \gamma_\mu \gamma_5 q ~,~
\label{eq:Leff}
\end{align}
the spin-dependent $\chi-$nucleon scattering cross section can be written as 
\begin{align}
\sigma^{p,n}_{\text{SD}} = \frac{12}{\pi} (\frac{m_\chi m_{p,n}}{m_\chi + m_{p,n}})^2 |a_{p,n}|^2 ~,~ \label{sigsd}
\end{align}
where 
\begin{align}
a_{p,n} = \sum_q d_q \Delta q_{p,n} ~.~
\end{align}
The factors $\Delta q_{p,n}$ parameterise the corresponding quark spin 
content of the nucleon:
\begin{align}
2 s_\mu \Delta q_{N} = <N| \bar{\psi}_q \gamma_\mu \gamma_5 \psi_q |N>~,~
\end{align}
where $s_\mu$ is the nucleon spin. The preferred values of 
the light quark contributions in the proton and neutron are:
\begin{align}
&\Delta_u^{(p)} = \Delta_d^{(n)} = 0.84, ~~ \Delta_d^{(p)} 
= \Delta_u^{(n)} = -0.43, ~~ \nonumber \\  
&\Delta_s^{(p)} = \Delta_s^{(n)} = -0.02~,~ 
\end{align}
where the strange quark contribution is motivated by a recent 
lattice QCD calculation~\cite{QCDSF:2011aa}.

\begin{figure}[htb]
\centering
 \includegraphics[width=0.12\textwidth]{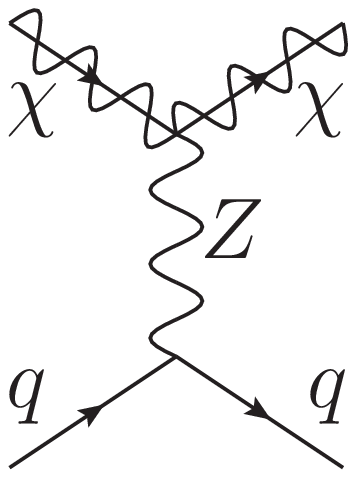} ~~~~~
  \includegraphics[width=0.15\textwidth]{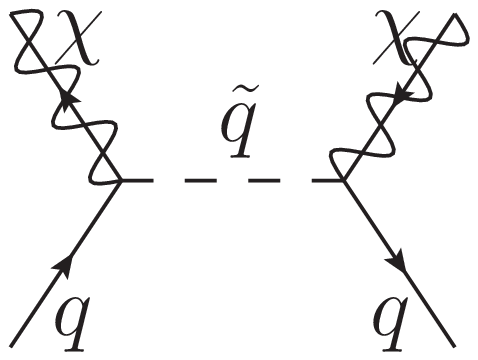}
 \caption{\label{prods} Processes that contribute to the Dark Matter 
spin dependent cross section for scattering from a nucleon. }
\end{figure}

In the MSSM at tree level, there are two processes which can contribute 
to the effective Lagrangian. 
The corresponding Feynman diagrams are given in Fig.~\ref{prods}. 
From those processes, we are able to calculate the coefficients of the 
effective Lagrangian from the renormalisable Lagrangian below:
\begin{align}
\mathcal{L} = \bar{q} (a_q +b_q \gamma_5) \chi \tilde{q} + c \bar{q} \gamma^\mu \gamma^5 q  Z_\mu + d \bar{\chi} \gamma^\mu  \gamma_5  \chi Z_\mu ~.~ \label{renorl}
\end{align}
The corresponding couplings in the MSSM are written as~\cite{Rosiek:1995kg} 
\begin{align}
a_u =& i \frac{Z^L_{\tilde{u}}}{2}(\frac{-g}{\sqrt{2} c_w}(\frac{1}{3} Z_N^{11} s_w + Z^{21}_N c_w) - Y_u Z^{41}_N   )  \nonumber \\ 
&+ i\frac{Z^R_{\tilde{u}}}{2} (\frac{2\sqrt{2} g s_w }{3 c_w} Z_N^{11} - Y_u Z^{41}_N )  \\
b_u =& i \frac{Z^L_{\tilde{u}}}{2}(\frac{g}{\sqrt{2} c_w}(\frac{1}{3} Z_N^{11} s_w  + Z^{21}_N c_w) - Y_u Z^{41}_N   )  \nonumber \\  
& + i\frac{Z^R_{\tilde{u}}}{2} (\frac{2\sqrt{2} g s_w }{3 c_w} Z_N^{11} + Y_u Z^{41}_N )  \\
a_d =& i \frac{Z^L_{\tilde{d}}}{2}(\frac{-g}{\sqrt{2} c_w} (\frac{1}{3} Z^{11}_N s_w -Z^{21}_N c_w) + Y_d Z^{31}_N)  \nonumber \\  
&+ i \frac{Z^R_{\tilde{d}}}{2} (\frac{ - \sqrt{2} g s_w}{3 c_w}Z^{11}_N + Y_d Z^{31}_N ) \\
b_d =& i \frac{Z^L_{\tilde{d}}}{2}(\frac{g}{\sqrt{2} c_w} (\frac{1}{3} Z^{11}_N s_w -Z^{21}_N c_w) + Y_d Z^{31}_N)  \nonumber \\  
&+ i \frac{Z^R_{\tilde{d}}}{2} (\frac{ - \sqrt{2} g s_w}{3 c_w}Z^{11}_N - Y_d Z^{31}_N )\\
c =& \frac{i}{2} \frac{g}{c_w} T_{3q} \\
d  =& - \frac{i }{4} \frac{g}{c_w} ((Z_N^{41})^2 - (Z_N^{31})^2)  
\end{align}

We consider first the $Z$ boson mediated amplitude in the 
non-relativistic limit:
\begin{align}
\mathcal{M}^Z_{\text{SD}} &=  c ~d~ \bar{\chi}  \gamma^\mu \gamma_5 \chi 
\frac{-i g_{\mu \nu}}{Q^2 - m_Z^2} \bar{q} \gamma^\nu \gamma_5 q \nonumber \\
   & \sim c~ d ~ \frac{i}{m_Z^2}(1+\mathcal{O}(m_Z^{-2}))  \bar{\chi}  
\gamma^\mu \gamma_5 \chi  \gamma_\mu \gamma_5 q \nonumber \\
  & \sim  c~d~ \frac{i}{m_Z^2} \bar{\chi}  \gamma^\mu \gamma_5 \chi  
\gamma_\mu \gamma_5 q \, ,
\end{align}
so that the effective coupling $d_q$ in Eq.~(\ref{eq:Leff}) is:
\begin{align}
d_q = \frac{cd}{m^2_Z} = \frac{g^2}{4 m_W^2} T_{3q} ((Z_N^{41})^2-(Z_N^{31})^2). \label{zcont}
\end{align}
Next, for the $\tilde{q}$ mediated process we find:
\begin{align}
&\mathcal{M}^{\tilde{q}}_{\text{SD}} = \bar{\chi} (a - b \gamma_5) q 
\frac{i}{(p_\chi + p_q)^2 - m_{\tilde{q}}^2} \bar{q} (a+b \gamma_5) \chi \nonumber \\
 &\sim \frac{-i}{m^2_{\tilde{q}} - (m_\chi + m_q)^2}  \bar{\chi} 
(a - b \gamma_5) q ~  \bar{q} (a+b \gamma_5) \chi \nonumber \\
 &= \frac{-i}{m^2_{\tilde{q}} - (m_\chi + m_q)^2} 
(a^2 \bar{\chi} q ~ \bar{q} \chi - b^2 \bar{\chi} \gamma_5 q ~ \bar{q} \gamma_5 \chi) \nonumber \\
 &\ni \frac{-i}{m^2_{\tilde{q}} - (m_\chi + m_q)^2} (\frac{a^2 + b^2}{4} 
\bar{\chi} \gamma^\mu \gamma_5 \chi ~ \bar{q} \gamma_\mu \gamma_5 q  ) \, ,
\end{align}
In this case the effective coupling in Eq.~(\ref{eq:Leff}) is:
\begin{align}
d_q = -\frac{1}{4} \frac{a^2 + b^2}{m^2_{\tilde{q}} - (m_\chi + m_q)^2} \label{qcont} ~.~
\end{align}
Note that the tree level effective coupling $d_q$ is only reliable when $m_{\tilde{q}} - m_{\chi}$  
is significantly larger than $m_q$. Some discussions regarding the precision of the tree level approximation are given in Appendix~\ref{app:prec}. 
And we have also checked that the result calculated from Eq.~(\ref{qcont}) matches well with numerical tool micrOMEGAs for light flavor quark. 

%%%%%%%%%%%%%%%%%%%%%%%%%%%%%%%%%%%%%%%%%%%%%%%%%%%%%%%%%%%%%%%%%%%%%%%%%%%%%%%%%%%%
\section{Light sbottom contribution}
\label{sec:sbsd}

For most processes of physical interest the Appelquist-Carrazone 
theorem tells us that heavy quark contributions are suppressed by 
order $1/m_Q^2$. However, as explained in the introduction, 
because of the $U(1)$ axial anomaly, 
the heavy quark contributions to spin-dependent quantities are only 
logarithmically suppressed. The particular case where this has been 
explored in great detail is the neutral weak charge of the proton.
Without heavy quarks this is just $\Delta u \, - \, \Delta d \, -
\, \Delta s$, which has been used to infer values of $\Delta s$. However, 
for a precise determination one must include the radiative corrections 
involving heavy quark loops which enter at order $1 / \ln m_Q$. For example, 
one finds a LO correction from the $b$-quark equal to~\cite{Bass:2002mv}:
\begin{align}
\label{eq:Delta-b}
\Delta^{(p)}_b = -\frac{6}{23 \pi} \tilde{\alpha}_b (\Delta^{(p)}_u +
\Delta^{(p)}_d +\Delta^{(p)}_s) \sim -0.0066 ~.~
\end{align}
We note that Eq.~(\ref{eq:Delta-b}) is second order in the strong coupling 
at the $b$ mass, as  is evident in the residual 5-flavor factor 
6/23 appearing there. However, the 
regularisation of the triangle diagram leads to a logarithm in $m_b$ in 
the numerator which has been used to cancel the logarithm in one 
factor of $\tilde{\alpha}_b$. 
The logarithmic radiative correction $\Delta^{(p)}_b$ is around 2 order of magnitude below the  $\Delta^{(p)}_u$. We will show later that with further enhancement from resonant effect the contribution from $\Delta^{(p)}_b$ can easily become dominant in the spin-dependent $\chi$-nucleon scattering. 

%This is the result obtained by all authors~\cite{Bass:2002mv,Bass:2005ku,Crewther:2005th,Bass:2002wy} who have 
%examined the problem since the work of Zee {\it et al.}\cite{Collins:1978wz}
%In summary, the b-loop yields a term that goes like $(\alpha_b)^2 \text{ln}(m_b)$, 
%which is indeed second order in the strong coupling but only suppressed 
%by one power of $\text{ln}(m_b)$.

Provided that the difference between the sbottom mass and that of the DM 
candidate is significantly larger than the mass 
of the $b$-quark, the $\tilde{q}$ propagator in Fig.~\ref{prods} 
can be effectively factored out, leaving the familiar triangle 
diagram which involves the $U(1)$ axial anomaly. In this case 
the bottom contribution to the axial charge of the target proton 
can be taken directly from Eq.~(\ref{eq:Delta-b}). 
We shall take the running coupling $\tilde{\alpha}_b = 0.2$. 
As a result, for the $Z$-mediated process, the contribution of 
$\Delta^{(p)}_b$ can only change the result by a factor of
\begin{align}
(\frac{\Delta^{(p)}_u -\Delta^{(p)}_d -\Delta^{(p)}_s - 
\Delta^{(p)}_b}{\Delta^{(p)}_u -\Delta^{(p)}_d -\Delta^{(p)}_s})^2 
\sim 1.01 \, ,
\end{align}
which is clearly very small.

\begin{figure}[htb]
\centering
 \includegraphics[width=0.4\textwidth]{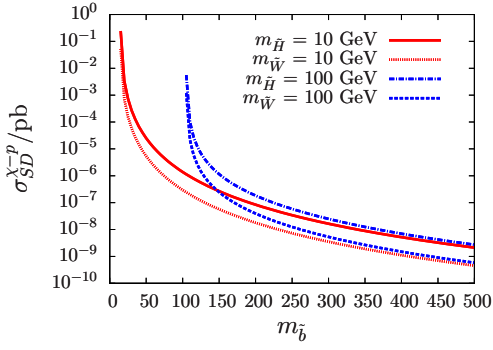}
 \caption{\label{sbsd} Sbottom contribution to the SD scattering cross 
section for Wino and Higgsino Dark Matter from the proton.  }
\end{figure}

On the other hand, the term involving $\Delta^{(p)}_b$ can give 
a significant contribution to the spin-dependent $\chi$-nucleon 
cross section when $\tilde{b}$ is relatively close in mass to the 
DM candidate, i.e. with resonant enhancement.  For simplicity we study the cases where the DM 
particle is either pure Wino or pure Higgsino.
The corresponding  couplings are:
\begin{align}
a^{(\tilde{W})}_b = \frac{i g}{2 \sqrt{2}} Z^L_{\tilde{b}}  , ~~~ & a^{(\tilde{H})}_b =  \frac{i}{2}Y_b Z^{31}_N (Z_{\tilde{b}}^L + Z_{\tilde{b}}^R)  \\
b^{(\tilde{W})}_b = -\frac{i g}{2 \sqrt{2}} Z^L_{\tilde{b}}, ~~~ & b^{(\tilde{H})}_b =  \frac{i}{2}Y_b Z^{31}_N (Z_{\tilde{b}}^L - Z_{\tilde{b}}^R)
\end{align}
where $Y_b=  \frac{g}{\sqrt{2} m_W \cos \beta} m_b$.
So, the cross section can be written as
\begin{align}
\sigma^{\tilde{b}- \tilde{W}}_\text{SD} & = \frac{12}{\pi} (\frac{m_\chi m_p}{m_\chi + m_p})^2 (-\frac{g^2 (T_{3b}  Z^L_{\tilde{b}})^2 }{4(m^2_{\tilde{b}} - (m_\chi + m_b)^2)} \Delta^{(p)}_b)^2  \label{sbwsd} \\
\sigma^{\tilde{b} - \tilde{H}}_\text{SD} &=  \frac{12}{\pi} (\frac{m_\chi m_p}{m_\chi + m_p})^2 (-\frac{0.5 Y_b^2 (Z_N^{31})^2}{4(m^2_{\tilde{b}} - (m_\chi + m_b)^2)}   \Delta^{(p)}_b)^2 \label{sbhsd}
\end{align}
where we have assumed the gauge eigenstate limit  
and only the sbottom mediated process is contributing. 
By fixing $m_\chi$ at either 10 or 100 GeV and taking $Z^L_{\tilde{b}} =1$ 
and  $\tan \beta=40$, $Z^{31}_N=\frac{1}{\sqrt{2}}$ for 
Wino and Higgsino DM, respectively, we can calculate the 
corresponding cross section as a function of $m_{\tilde{b}}$. 
The result is shown in Fig.~\ref{sbsd}. 
From Fig.~\ref{sbsd} we see that the sbottom can give a very large 
contribution when the mass splitting 
$m_{\tilde{b}} - m_\chi$ is $ \lesssim 100$ GeV. 

\subsection{Comparison with the contribution from the first generation squarks}
First, we study the simpler case where the DM is gaugino. In this case  
there is no coupling between the $Z$ boson and DM and 
only the squark mediated process can contribute to the SD interaction. 
In this subsection, we investigate the extent to which 
the sbottom should be lighter than first generation squark, 
so that they at least have comparable cross sections. 
In the following we consider 
the sum of the contributions of all first generation 
squarks ($\tilde{u}_L, \tilde{u}_R, \tilde{d}_L, \tilde{d}_R$), 
with their masses taken to be degenerate for simplicity.

Assuming that the DM is either pure Wino or Bino, the ratio of 
the corresponding SD cross section for sbottom to the sum of the 
contributions from all the first generation squarks can be calculated as
\begin{align}
& \frac{\sigma_{\text{SD}}^{\tilde{b}-\tilde{W}}}{\sigma_{\text{SD}}^{\tilde{q}_{u,d}-\tilde{H}} } =  \frac{(Z^L_{\tilde{b}})^2 |\Delta^{(p)}_b|}{m^2_{\tilde{b}} - (m_\chi + m_b)^2} / \frac{\Delta^{(p)}_u + \Delta^{(p)}_d }{m^2_{\tilde{q}_{u,d}} - m^2_\chi} \\
&\frac{\sigma_{\text{SD}}^{\tilde{b}-\tilde{B}}}{\sigma_{\text{SD}}^{\tilde{q}_{u,d}-\tilde{B}} } = 
\frac{(a^2 +b^2)_{\tilde{b}_{1}}}{m^2_{\tilde{b}} - (m_\chi + m_b)^2} \times \nonumber \\
&(\frac{ \sum_{\tilde{u}_{L,R}} (a^2 +b^2)  \Delta^{(p)}_u 
+ \sum_{\tilde{d}_{L,R}} (a^2+b^2) 
\Delta^{(p)}_d }{m^2_{\tilde{q}_{u,d}} - m^2_\chi})^{-1} \, .
\end{align}
 \begin{figure}[htb]
\centering
  \includegraphics[width=0.35\textwidth]{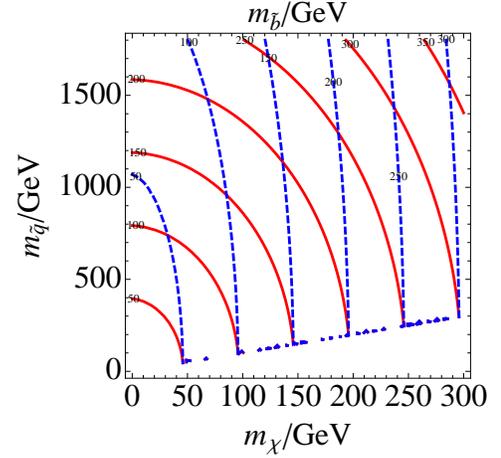}
 \caption{\label{cont} Contours of constant $m_{\tilde{b}}$ which show 
where the sbottom and degenerate first 
generation squarks give the same contribution for Wino (Red solid line) 
and Bino (Blue dashed line) DM. }
\end{figure}
The corresponding contours of ${\sigma_{\text{SD}}^{\tilde{b}-\chi}} 
= {\sigma_{\text{SD}}^{\tilde{q}_{u,d}-\chi} }$ are shown in Fig.~\ref{cont}. 
The case of Wino DM is more interesting than that of Bino DM because 
of its larger $g_2$ coupling.  In this case, for 1.5 TeV first generation 
squarks and $\mathcal{O}(100)$ GeV DM, an sbottom lighter than 
about 200 GeV gives a larger cross section than the first generation squarks. 
On the other hand, for Bino DM, a much lighter sbottom($\sim 110$ GeV) 
is required -- too light for the present calculation to be reliable. 

\subsection{Contribution coherent with that of the $Z$-boson}
When the DM is predominantly Higgsino, the first two generation squark 
mediated processes are greatly suppressed by their small Yukawa couplings. 
Its couplings to the Z boson and sbottom are dependent on the mixing 
between the two Higgsino states. 

Firstly, we briefly discuss the Higgsino mixing in the MSSM. 
In the basis ($\tilde{B}, \tilde{W},\tilde{H}^0_d , \tilde{H}^0_u$), 
the neutralino mass matrix is given by:
\begin{align}
M_{N}  =  \left(
\begin{array}{cccc}
M_1 & 0 & -c_\beta s_W m_Z & s_\beta s_W m_Z \\
0 & M_2 & c_\beta c_W m_Z & - s_\beta c_W m_Z \\
-c_\beta s_W m_Z & c_\beta c_W m_Z & 0 & -\mu \\
s_\beta s_W m_Z & -s_\beta c_W m_Z & -\mu & 0
\end{array}
\right)
\end{align}
{}From the mass matrix we conclude that if
\begin{align}
m_Z \ll \mu, M_1, M_2~,~
\end{align}
the four neutralino mass eigenstates $\tilde{N}_i$ will be Bino $\tilde{B}$ dominated, Wino $\tilde{W}$ dominated and Higgsino $(\tilde{H}^0_u \pm \tilde{H}^0_d)/\sqrt{2}$ dominated, respectively. 
For example, if we also decouple the Bino and Wino from the mass matrix, 
the component difference between $\tilde{H}^0_u$ and $\tilde{H}^0_d$, 
{}for a given mass eigenstate is 
\begin{align}
\Delta N_{\tilde{H}^0_u - \tilde{H}^0_d} \propto \frac{m_Z^2}{M_i \mu}
\end{align}
For a few TeV gaugino and a few hundred GeV Higgsino, 
$\Delta N_{\tilde{H}^0_u - \tilde{H}^0_d} \sim \mathcal{O}(10^{-2})$. 
Then, the contribution from the $Z$ boson mediated process can be estimated by
\begin{align}
\sigma^Z_{\text{SD}} &= \frac{12}{\pi} (\frac{m_\chi m_p}{m_\chi+m_p})^2 (\sum_{q=u,d,s} d_q \Delta q^{(p)})^2 \\
 &= \frac{12}{\pi} (\frac{m_\chi m_p}{m_\chi+m_p})^2 ( 
\frac{g^2}{8 m^2_W}( (Z^{41}_N)^2-(Z^{31}_N)^2) )^2   \nonumber \\  
& (\Delta^{(p)}_u \times 
\frac{1}{2} + \Delta^{(p)}_d \times 
(- \frac{1}{2}) + \Delta^{(p)}_s \times( - \frac{1}{2} ))^2 ~,~ 
\label{zval}
\end{align}
which is $\sim 10^{-6}$ pb. From Fig.~\ref{sbsd}, we conclude that 
this corresponds to $m_{\tilde{b}} \sim 150$ GeV for $m_\chi \sim 100$ GeV. 

To have a closer look at the coherent effects of $Z$ boson and sbottom 
mediated processes, we have chosen the decoupled Wino/Bino limit, 
with the Higgsino DM mixing:
\begin{align}
\chi = a \tilde{H}_d + b \tilde{H}_u ~,~
\end{align}
where $a^2 + b^2 =1$ and $b=1.01 a$, as argued previously.  
This corresponds to a cross section for the $Z$ mediated process  
of order $\sim 3 \times 10^{-6}$ pb. 

In this region, the $\tilde{b}$ mediated process may also give 
a competitive contribution. 
As a result, the DM will have opposite sign coherent effects for the 
proton and neutron:
\begin{align} \label{sigsdz}
&\sigma_{\text{SD}}^{\chi - (p,n)}  = \frac{12}{\pi} (\frac{m_\chi m_p}{m_\chi + m_p})^2  
  ( ( \frac{g^2 ((Z_N^{41})^2 -(Z_N^{31})^2)}{8 m^2_W}) (T_{3u} \Delta_u^{(p,n)} \nonumber \\ 
  & + T_{3d} \Delta_d^{(p,n)} + T_{3s} \Delta_s^{(p,n)} )   - \frac{0.5 Y_b^2 (Z_N^{31})^2}{4(m^2_{\tilde{b}} - (m_\chi + m_b)^2 )} 
\Delta_b^{(p,n)}  )^2 \, .
\end{align}
This makes the detailed consequences for SD DM scattering from real 
nuclei~\cite{Vietze:2014vsa} potentially very complex.

\begin{figure}[htb]
\centering
 \includegraphics[width=0.4\textwidth]{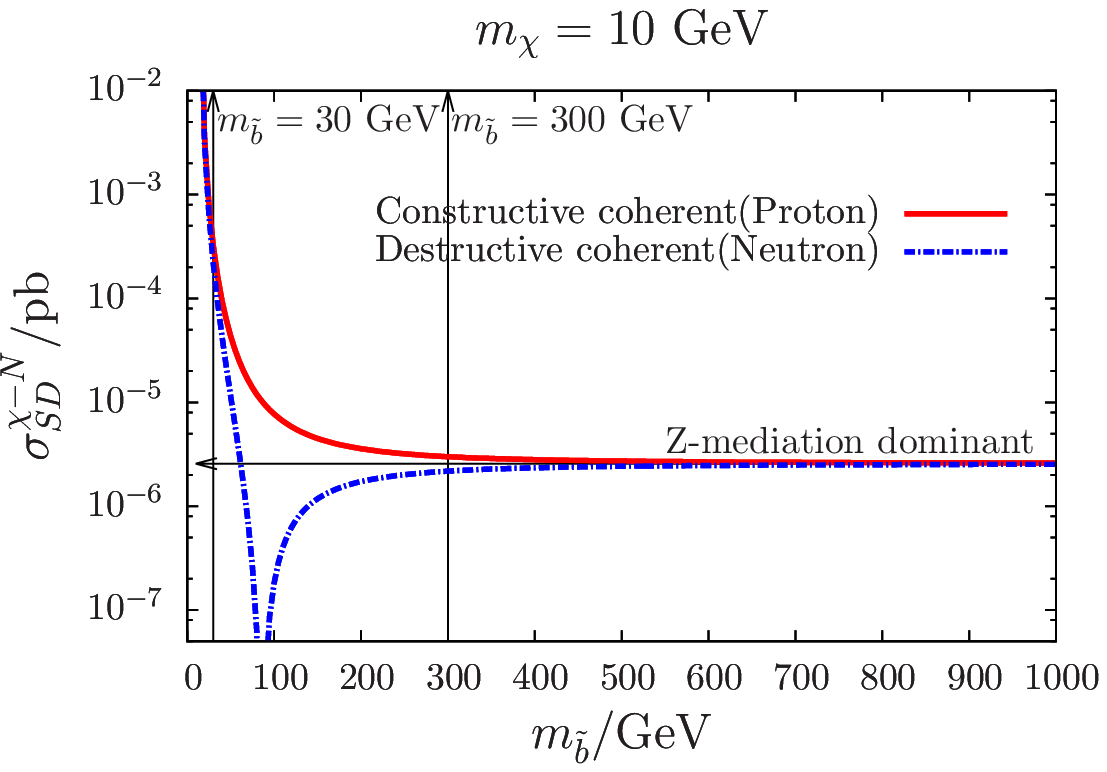}
  \includegraphics[width=0.4\textwidth]{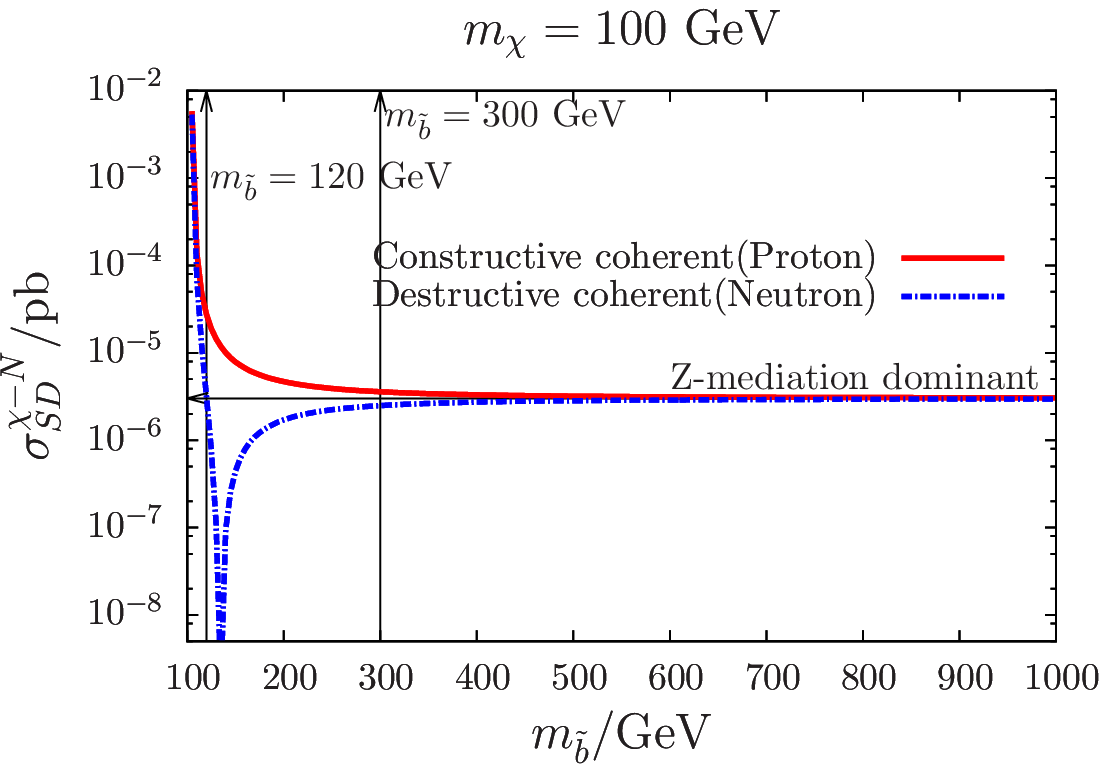}
 \caption{\label{combzb} Constructive and destructive coherence effects 
between $Z$ mediated and $\tilde{b}$ mediated processes for 
$m_\chi=10$ GeV (Upper) and $m_\chi=100$ GeV (Lower), respectively. The vertical arrowed lines  at $m_{\tilde{b}}$ = 30 GeV and $m_{\tilde{b}}$ = 120 GeV 
indicate the value below which the 
calculation should not be considered reliable.}
\end{figure}

We show the importance of the $b$-quark contribution through 
its coherent effects between $Z$ mediated and sbottom mediated processes 
in Fig.~\ref{combzb}.  
There we have marked out the compressed spectrum region 
($\Delta(m_{\tilde{b}} - m_\chi) \lesssim 20$ GeV) where our 
tree level calculation cannot be considered reliable. 
According to the Eq.~(\ref{sigsdz}) with assumed $Z_N^{41}=1.01 Z_N^{31}$, the first term in the parenthesis is positive for proton and negative for neutron, while the second term is always negative because $\Delta_b^{(p)}=\Delta_b^{(n)}<0$. As a result, the interference terms for proton and neutron are constructive and destructive, respectively.
We find that for the DM mass around  $\mathcal{O}(10-100)$ GeV an sbottom  
with mass $\lesssim 300$ GeV can make a non-negligible contribution. 
In some specific regions, the corresponding cross section for 
the Z-mediated process may even be enhanced  
or reduced by several orders of magnitude. 

\section{Spin-independent DM detection constraint from LUX}
\label{sec:si}
The same process shown in the right panel of Fig.~\ref{prods}, 
which can give rise to an enhancement of the spin-dependent scattering 
cross section, can also contribute to spin-independent scattering.  
As a result, the very stringent spin-independent DM search 
bound from LUX~\cite{Akerib:2015rjg} may already exclude some of 
the parameter region found to be of interest here. 

We start with the following effective 
Lagrangian~\cite{Griest:1988yr,Griest:1988ma,Drees:1993bu,Hisano:2010ct}:
\begin{align}
 \mathcal{L}_{\text{SI}} = &  \sum_q (f_q m_q \bar{\chi} \chi \bar{q} q + 
\frac{g^{(1)}_q}{m_\chi} \bar{\chi} i \partial^\mu \gamma^\nu 
\chi \mathcal{O}^q_{\mu \nu} \nonumber \\ 
&+ \frac{g^{(2)}_q}{m_\chi^2} \bar{\chi} 
(i \partial^\mu) (i \partial^\nu) \chi \mathcal{O}^q_{\mu \nu} ) 
  + f_G \bar{\chi} \chi G^a_{\mu \nu} G^{a \mu \nu}
\end{align}
where $\chi$ is the DM field, $m_\chi$ its mass and the twist-2 operator:
\begin{align}
\mathcal{O}^q_{\mu \nu} = \frac{1}{2} \bar{q} i (D_\mu \gamma_\nu + D_\nu \gamma_\mu -\frac{1}{2} g_{\mu \nu} \slashed{D}) q ~.~
\end{align}
The corresponding spin-independent scattering cross section of DM with 
a proton can be written as
\begin{align}
\sigma^{\chi-p}_{\text{SI}} = \frac{4}{\pi} \mu^2 (f_N)^2
\end{align}
where $\mu = m_\chi m_N / (m_\chi+ m_N)$ and 
\begin{align}
 \frac{f_N}{m_N} &= \sum_{q=u,d,s} f_q f_{T_q} + \sum_{q=u,d,s,c,b} \frac{3}{4} (q(2)+\bar{q}(2)) (g^{(1)}_q + g^{(2)}_q)  \nonumber \\ 
 &- \frac{8 \pi}{ 9 \alpha_s} f_{T_G} f_G  \label{gg} \\
    & \sim  \sum_{q=u,d,s} f_q f_{T_q} + \sum_{q=u,d,s,c,b} \frac{3}{4} (q(2)+\bar{q}(2)) (g^{(1)}_q + g^{(2)}_q)  \nonumber \\ 
    &+  \frac{2}{27}\sum_{Q=c,b,t} f_{T_G} f_{Q} \label{qq}
\end{align}
The light quark parameters $f_{T_q}$ are defined by
\begin{align}
f_{T_q} m_N = <N| m_q \bar{q} q |N>~,~
\end{align}
and $f_{T_G} = 1- \sum_{q=u,d,s} f_{T_q}$.
Recent lattice simulations 
give~\cite{Ohki:2008ff,Giedt:2009mr,Young:2009zb,Shanahan:2012wh}:
\begin{align}
f^p_u = 0.023, ~~ f^p_d = 0.033, ~~ f^p_s = 0.026 ~.~
\end{align}
The second moments of the parton distribution functions(PDFs) 
can be used to evaluate the matrix element of $\mathcal{O}^q_{\mu \nu}$:
\begin{align}
(p_\mu p_\nu - \frac{1}{4} m^2_N g_{\mu \nu} ) (q(2) + \bar{q}(2)) = 
m_N <N(p)| \mathcal{O}^q_{\mu \nu}| N(p)> \, ,
\end{align}
which from the CTEQ PDFs~\cite{Pumplin:2002vw} yields 
\begin{align}
b(2)=0.012, ~~~  \bar{b}(2)=0.012 ~,~
\end{align}
at the $Z$ boson mass scale.

\begin{comment}
The full gluon contribution at NLO in Eq.~(\ref{gg}) for the SI process 
has been calculated in Ref.~\cite{Drees:1993bu,Hisano:2010ct}. For 
the light quarks, the gluon contribution from Fig.~\ref{l1} is 
much smaller than the tree level contribution, especially 
when $m_\chi$ is non-negligible. 
Other Feynman diagrams vanish in the Fock-Schwinger gauge. 
As for heavy quarks, the gluon contribution is dominated by Fig.~\ref{l2} 
in this limit, even though Fig.~\ref{l1} is important when $m_\chi$ 
is small. Ref.~\cite{Drees:1993bu} pointed out that the result of 
the NLO calculation in Fig.~\ref{l2} is the same as the 
tree level calculation in Fig.~\ref{t1}, using Eq.~(\ref{qq}) to 
the leading order of $m_{\tilde{q}}^{-2}$.
%
\begin{figure}
        \centering
        \subfigure[Tree level]{
                \includegraphics[width=0.15\textwidth]{t1}
                \label{t1}}
        ~ 
        \subfigure[Short-distance Contribution]{
                \includegraphics[width=0.12\textwidth]{l1}
                \label{l1}}
        ~ 
        \subfigure[Long-distance Contribution]{
                \includegraphics[width=0.12\textwidth]{l2}
                \label{l2}}
        \caption{Spin-independent $\chi - $nucleon scattering 
processes. }
\label{feyns}
\end{figure}
%
\end{comment}
Using a similar technique to that used in calculating the SD effective 
coefficient, $d_q$, above, we can find the corresponding effective 
coefficient for the spin-independent case.  
Based on  the renormalizable Lagrangian Eq.~(\ref{renorl}), we have
\begin{align}
f_q &= \frac{m_\chi}{(m^2_{\tilde{q}} - (m_\chi+m_q)^2)^2} 
\frac{a_q^2 + b^2_q}{8}  \nonumber \\ 
&- \frac{1}{m_q (m^2_{\tilde{q}} - 
(m_\chi+m_q)^2)} \frac{a^2_q - b^2_q}{4} \\
g^{(1)}_q + g^{(2)}_q &= \frac{m_\chi}{(m^2_{\tilde{q}} - 
(m_\chi+m_q)^2)^2} \frac{a^2_q + b^2_q}{2} \, .
\end{align}
As a result the twist-2 operator, 
$\mathcal{O}^q_{\mu \nu}$, gives a much larger contribution than $f_q$ 
in most cases. For Higgsino DM, where $a_q \neq b_q$, the second 
term of $f_q$ can easily become dominant. However in this case it is 
negative, so a cancellation between $f_q$ and $g_q$ may happen 
in some of the parameter regions.  

\begin{figure}[htb]
\centering
 \includegraphics[width=0.4\textwidth]{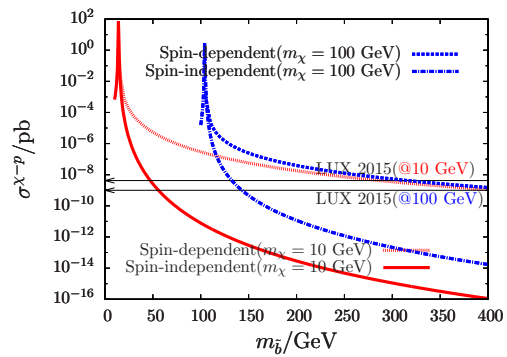}
 \caption{\label{bwsi} SD and SI scattering cross 
section for Wino DM from the proton,
with $m_\chi=10$ GeV and $m_\chi=100$ GeV, respectively. }
\end{figure}

We first consider the pure Wino DM case, with only $\tilde{b}_L$ 
mediated scattering. From Eq.~(\ref{qq}), we have
\begin{align}
f_N = m_p (\frac{3}{4} (b(2) + \bar{b}(2)) (g_b^{\tilde{b}_L -\tilde{W}}) + \frac{2}{27} f_{T_G}f_b^{\tilde{b}_L -\tilde{W}}) \label{bwsifn}
\end{align}
where 
\begin{align}
f_b^{\tilde{b}_L -\tilde{W}} &= \frac{g^2 m_\chi}{32} \frac{1}{(m^2_{\tilde{b}} - m^2_\chi)^2}\\
g_b^{\tilde{b}_L -\tilde{W}} &= \frac{g^2 m_\chi}{8} \frac{1}{(m^2_{\tilde{b}} - m^2_\chi)^2} 
\end{align}
The  tree level calculation for the spin-independent and spin-dependent 
cross sections is shown in Fig.~\ref{bwsi}.  
We conclude from the figure that $m_{\tilde{b}} - m_\chi \gtrsim 50$ GeV 
is required to evade the spin-independent bound from LUX for Wino DM.
It has to be noted that the pole at $m_{\tilde{b}} = m_b + m_\chi $,
{}for SI tree level results, will not show up when the full NLO effects 
are taken into account~\cite{Gondolo:2013wwa}. 
We have checked that our results fit the numerical results from 
micrOMEGAs~\cite{Belanger:2008sj,Belanger:2013oya}
quite well, outside the pole region. 

Next, we discuss the more interesting case where the DM is predominantly 
Higgsino. As discussed above, in this case the relatively large spin 
dependent cross section from the sbottom mediated process can 
interfere coherently with the $Z$ mediated process, leading to very 
different SD scattering rates for protons and neutrons. 

The SI DM-proton effective coupling is
\begin{align}
f_N = m_p (\frac{3}{4} (b(2) + \bar{b}(2)) (g_b^{\tilde{b}_1 -\tilde{H}}) + \frac{2}{27} f_{T_G}f_b^{\tilde{b}_1 -\tilde{H}}) \label{bhsifn}
\end{align}
where 
\begin{align}
f_b^{\tilde{b}_1 -\tilde{H}}  &= \frac{m_\chi}{(m_{\tilde{b}}^2- (m_\chi + m_b)^2 )^2} \frac{0.5 Y_b^2 (Z_N^{31})^2 }{8}  \nonumber \\ 
&- \frac{1}{m_b (m_{\tilde{b}}^2 - (m_\chi + m_b)^2)} \frac{ Y_b^2 (Z_N^{31})^2 Z^L_{\tilde{b}} Z^R_{\tilde{b}}}{4} \\
g_b^{\tilde{b}_1 -\tilde{H}} &= \frac{m_\chi}{(m_{\tilde{b}}^2 - 
(m_\chi + m_b)^2)^2} \frac{0.5 Y_b^2 (Z_N^{31})^2}{2} \, .
\end{align}
\begin{figure}[h!]
\centering
 \includegraphics[width=0.4\textwidth]{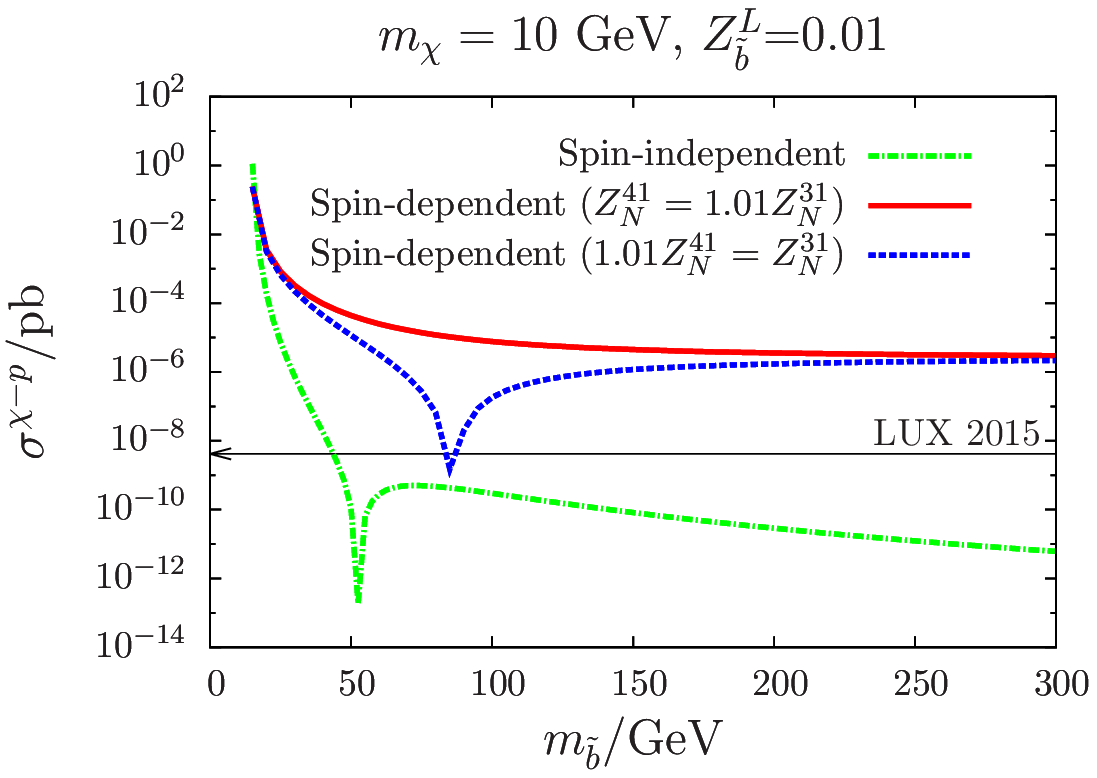} 
  \includegraphics[width=0.4\textwidth]{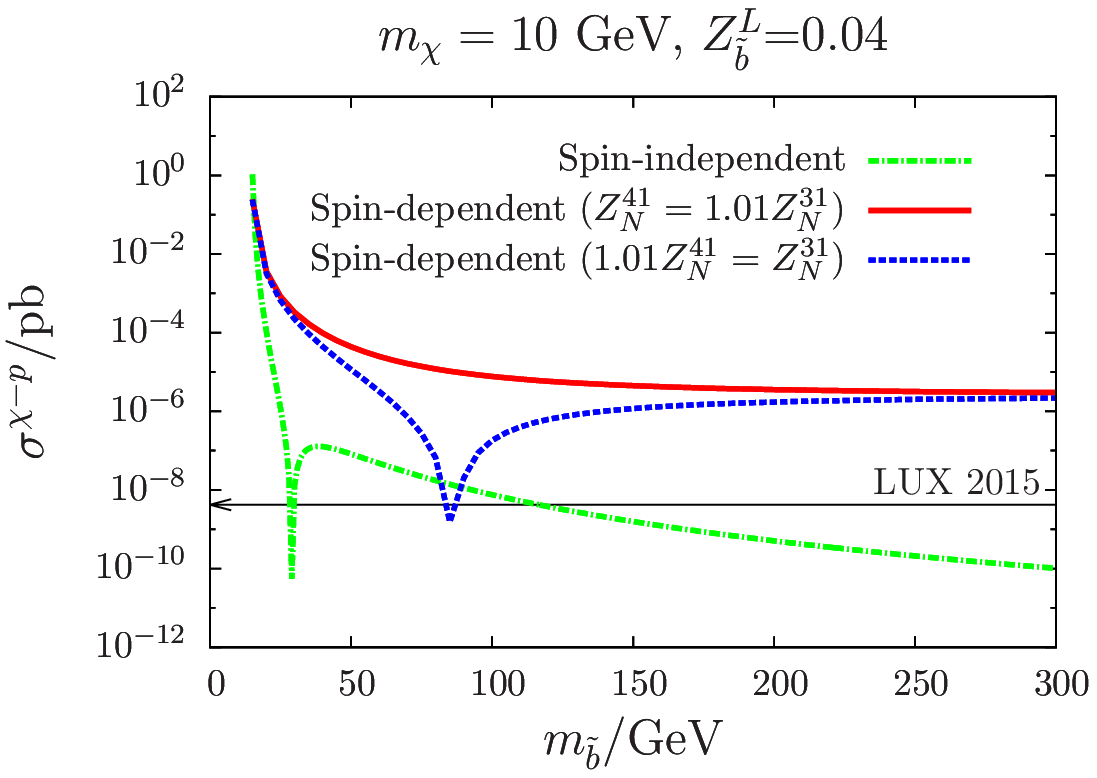} 
 \caption{\label{bhcomp}  SD and SI scattering cross 
section for different mixing of Higgsino DM from the proton.
Upper: $m_\chi=10$ GeV, $Z^L_{\tilde{b}}=0.01$. Lower: $m_\chi=10$ GeV, $Z^L_{\tilde{b}}=0.04$.  
}
\end{figure}
The corresponding tree level calculation for the spin-dependent and 
spin-independent cross sections  is shown in Fig.~\ref{bhcomp}. 
From that figure we see that a small component of left-handed 
sbottom is favoured in order to evade the LUX bound.
When the left-handed sbottom component is 
relatively large, a large SD cross section may also be consistent with 
the LUX experiment if there is a cancellation in $\sigma_{\text{SI}}$. 

\section{Model dependent constraints and a general argument}
\label{sec:cons}
We have presented a representative study of the potential importance of 
the bottom quark contribution to DM spin-dependent detection within 
the framework of the MSSM. This particular contribution has hitherto 
been overlooked. However, in a realistic model such as MSSM, 
there are many other incidental constraints.  We will briefly outline 
how these may be evaded, while keeping our discussion 
as general as possible in this section. 

LEP placed a very stringent bound on the chargino 
mass ($m_{\tilde{H}^\pm (\tilde{W}^\pm)} > 92.4(91.9)$ GeV)~\cite{Heister:2002mn}. 
Because for either Wino and Higgsino DM there is a charged partner (chargino), 
which has very similar mass with the DM, we cannot have Wino and Higgsino 
DM of $m_\chi \lesssim 90$ GeV in a typical MSSM framework.

\begin{figure}[htb]
\centering
 \includegraphics[width=0.4\textwidth]{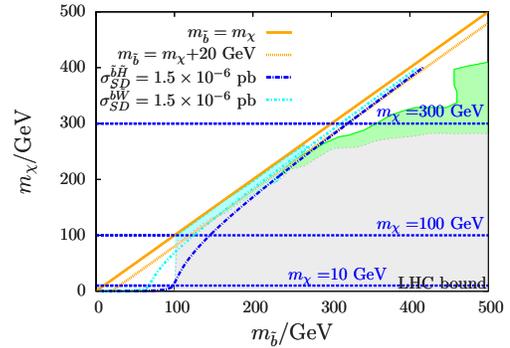}
 \caption{\label{lhc}  LHC exclusion bound on sbottom mass
versus DM mass. The grey, green and cyan shaded region correspond to the exclusion limits given by LHC searches for sbottom at 8 TeV~\cite{Aad:2013ija}, 13 TeV~\cite{Aad:2013ija} and searches for mono-jet at 8 TeV~\cite{Aad:2014nra}, respectively.}
\end{figure}

As for $m_\chi \gtrsim 100$ GeV, on the other hand, it will be 
constrained  by LHC sbottom searches~\cite{Aad:2013ija,ATLAS-CONF-2015-066} and mono-jet search~\cite{Aad:2014nra}, 
since we usually need a relatively light sbottom to enhance the bottom 
quark spin dependent contribution. The corresponding LHC exclusion 
bounds and spin-dependent $\chi -p$ scattering cross section in 
the $m_{\tilde{b}} -m_\chi $ plane are shown in Fig.~\ref{lhc}. 
To generate this figure we have used Eq.~(\ref{sbwsd}) and~(\ref{sbhsd}), 
where only the sbottom mediated process is considered. The contours of 
$\sigma_{\text{SD}}^{\chi-p}$ show the condition when the sbottom 
mediated contribution is half the size of Z mediated process for Wino DM 
and a typical Higgsino DM candidate with $Z_{N}^{41} = 1.01 Z_{N}^{31}$.  
This figure suggests that a large portion of parameter space is excluded by 
the LHC sbottom searches. 

However, there are several ways to avoid these constraints:
\begin{itemize}
\item For $m_\chi \lesssim 100$ GeV, we can work in a more general framework, 
where the dark matter does not have any charged partners. Its couplings 
to the $Z$ boson and $\tilde{b}$ may be of the same order; 
e.g. the simplified model framework~\cite{Abdallah:2014hon}  or 
{}flavored dark matter models~\cite{Batell:2011tc,Agrawal:2011ze}. 
\item For $m_\chi \lesssim 100$ GeV, if the charged Higgsino decays 
into DM and a relatively long lived particle, with lifetime 
of $\mathcal{O}(10-100)$ cm, similar to Ref.~\cite{Batell:2013bka}. 
As a result, the reconstructed track will not point to the interaction point 
and would therefore be unlikely to be considered a "good" track. 
In this case, the LEP constraints on charginos can be evaded.  
The light sbottom constraint can also be evaded by tuning 
appropriate mixing -- see e.g. Refs.~\cite{Batell:2013psa,Han:2014nba}.
\item For $m_\chi \gtrsim 100$ GeV, if the sbottom is decayed in more complicate modes  other than $\tilde{b} \to b \chi$,  the corresponding  LHC bound on sbottom mass can be loosened.
\item  We can also work with heavier DM, e.g. $m_\chi=300$ GeV, for example, 
as shown in Fig.~\ref{combzb300}. In this case, $m_{\tilde{b}} \lesssim 350$ GeV 
is consistent with the LHC searches, while the sbottom mediated process 
can give a significant contribution to spin-dependent scattering. 
 \begin{figure}[htb]
\centering
 \includegraphics[width=0.47\textwidth]{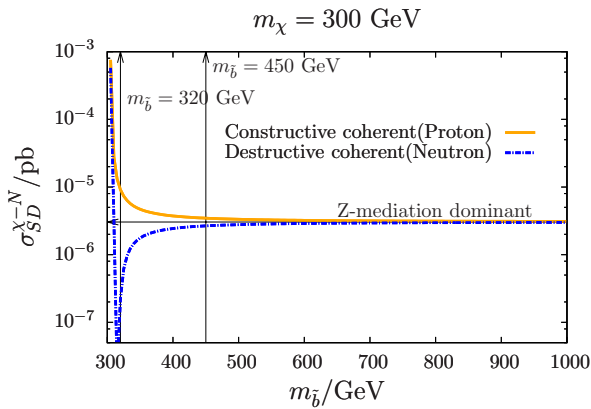}
 \caption{\label{combzb300} Constructive and destructive interference 
effects between $Z$ mediated and $\tilde{b}$ mediated processes for 
Higgsino DM with $m_\chi=300$ GeV.}
\end{figure}
\end{itemize}

%%%%%%%%%%%%%%%%%%%%%%%%%%%%%%
\section{Conclusion}
\label{sec:conc}
In this work, we have demonstrated the potential 
importance of the bottom quark 
contribution to the DM spin-dependent cross section 
due to the axial anomaly and resonant enhancement, 
which has hitherto been  overlooked. 
Even though our calculation was carried out within the framework of the MSSM, 
the general conclusion will be relevant to  any  models with 
similar particle content, since the only relevant ingredients 
are $\chi$, $\tilde{b}$ and the $Z$-boson. 

In the MSSM, we calculated the bottom quark contribution to 
spin-dependent $\chi-p$ scattering. Firstly, we considered Gaugino DM, 
where there is no coupling between the Z-boson and DM. 
Assuming $m_\chi=100$ GeV and degenerate first generation squarks 
with a  mass of 1.5 TeV, we found that an sbottom of 
mass $m_{\tilde{b}} \lesssim 200$ GeV can give rise to 
a larger spin-dependent cross section for Wino DM. By contrast, for 
Bino DM a much lighter sbottom mass ($m_{\tilde{b}}\sim110$ GeV) is 
required to give a competitive cross section.  As for Higgsino DM, 
the first generation squark contributions are suppressed by their small 
Yukawa couplings. However, the Z-boson mediated process does contribute. 
{}For a given Higgsino mixing of DM the sbottom mediated process may 
interfere either constructively or destructively 
with $Z$-boson mediated processes,
with different signs for protons and neutrons.  For a typical 
mixing of Higgsino DM with mass around $\mathcal{O}(10-100)$ GeV, we find that an 
sbottom of mass below $300$ GeV can have non-negligible effects.

The squark mediated process that gives rise to an increase 
in spin-dependent DM scattering can 
also contribute to the spin-independent cross section. Our calculation 
shows that $\Delta(m_{\tilde{b}} -m_{\chi}) \gtrsim 50$ GeV is 
required to evade the LUX constraint for Wino DM, while for Higgsino DM,  
either a small component of left-handed sbottom or a large cancellation 
in $\sigma_{\text{SI}}$ is needed. 
Some incidental model dependent constraints from LEP and the LHC are considered as well. Those constraints, however, can be evaded in more general theoretical frameworks.

As pointed out earlier, our tree level results may break down  
as the sbottom and DM masses become degenerate 
($\Delta(m_{\tilde{b}}-m_\chi ) \lesssim 20$ GeV). We leave 
the higher order calculation for this small region for  future work. 
Finally, while the calculations for the top quark case will be more 
complicated because there is no clear separation of mass scales for 
interesting ranges of DM mass, there is a clear need to investigate 
the role of the axial anomaly for that case too. 

\section*{Acknowledgements} 
This work was supported by the Australian Research Council through the 
ARC Centre of Excellence for Particle Physics at the Terascale 
(Grant CE110001004) and by an ARC Australian Laureate Fellowship 
(Grant FL0992247, AWT).

\appendix
\section{Precision of tree level approximation}
\label{app:prec}

\begin{figure}[h!]
\centering
 \includegraphics[width=0.4\textwidth]{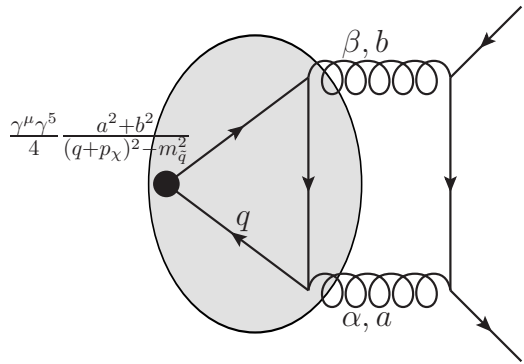} 
 \caption{\label{fig:2loop} Bottom quark contribution to the axial current. 
}
\end{figure}

The heavy quark contributions to the axial charge start at two loop level through the process shown in Fig.~\ref{fig:2loop}. A detailed calculation of this diagram is given in Ref.~\cite{Bass:2005ku,Crewther:2005th}. In this study, all we need to know is the $m_{\tilde{q}}$  dependence of the amplitude. Then we can derive the range of  $m_{\tilde{q}}$ for which the tree level effective coupling Eq.~\ref{qcont} is justified. 

\begin{figure}[htb]
\centering
 \includegraphics[width=0.4\textwidth]{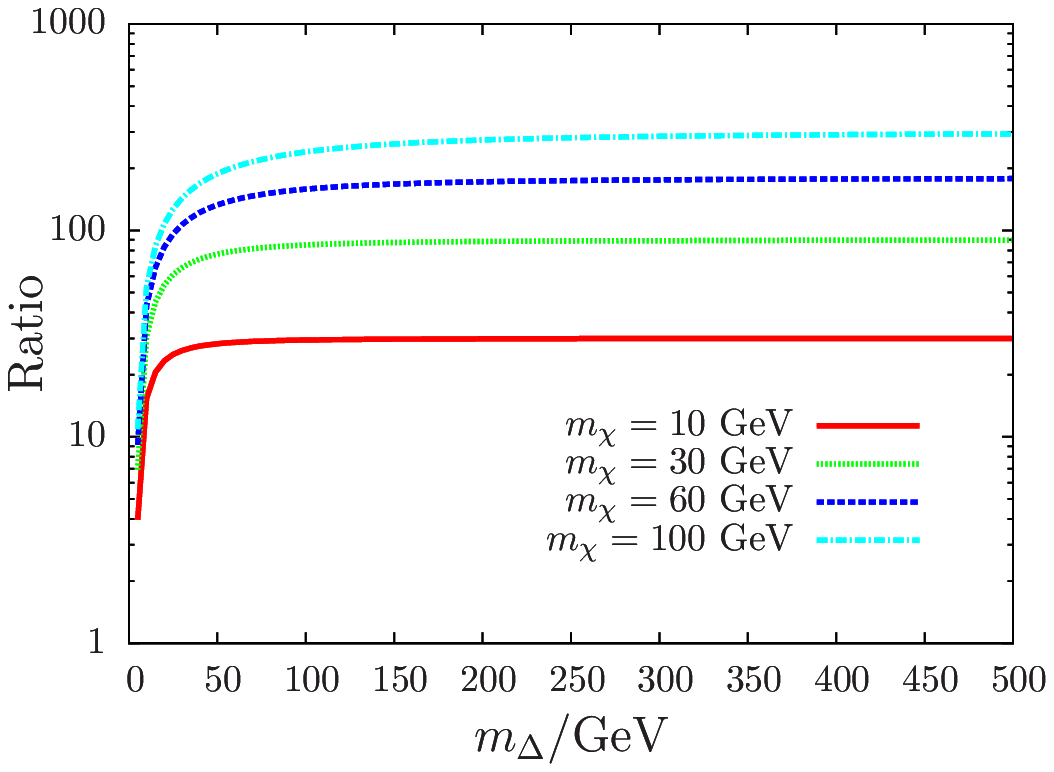} 
  \includegraphics[width=0.4\textwidth]{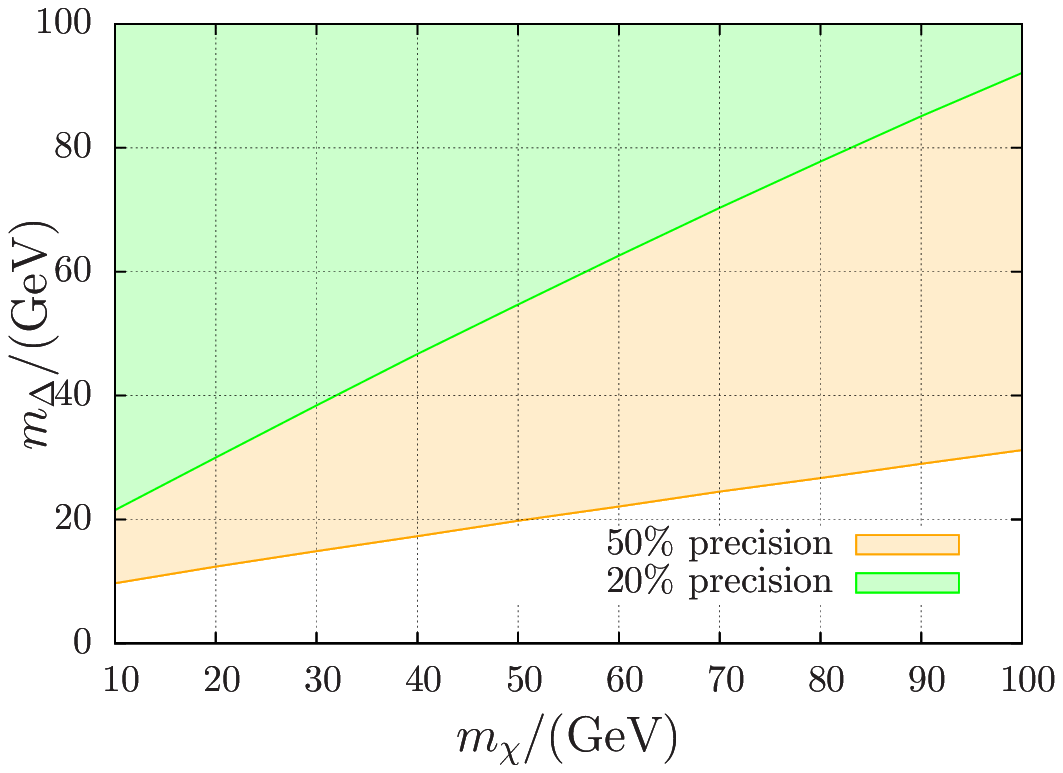}  
 \caption{\label{fig:prec} Upper: The $m_{\Delta}$ dependence of the ratio for several given DM masses.  Lower: In the shaded region, the tree level approximation matches the loop level results within required precision. 
}
\end{figure}

The $m_{\tilde{q}}$ dependence only exists in the triangle loop that is marked by the grey shaded ellipse in Fig.~\ref{fig:2loop}. The vertex amplitude is
\begin{align}
\Gamma^{a b}_{\mu \alpha \beta} &= \frac{a^2+b^2}{4} (-1) g^2 \int \frac{d^4 q}{(2 \pi)^4} \text{Tr} \{ \frac{\gamma_\mu \gamma_5}{ (q+p_\chi)^2-m^2_{\tilde{q}} } \cdot \nonumber \\
 &  \frac{i}{\slashed{q}-m} (i \gamma_{\alpha} \frac{1}{2} \lambda^a ) \frac{i}{\slashed{q}-m} (i \gamma_{\beta} \frac{1}{2} \lambda^b ) \frac{i}{\slashed{q}-m} \} ~,~\label{amploop1}
\end{align}
where $m,q$  is the quark mass and momentum in the triangle, $p_\chi$ is the dark matter four momentum and $m_{\tilde{q}}$ is the squark mass. 
After introducing Feynman parameter $x$ and the substitution:
\begin{align}
\l^\mu &= q^\mu +x p_\chi^\mu ~,~\\
\Delta &= x^2 p_\chi^2 - x p_\chi^2 +m^2(1-x) +x m^2_{\tilde{q}}  ~,~
\end{align}
the vertex amplitude Eq.~\ref{amploop1} can be simplified to 
\begin{align}
& \Gamma^{a b}_{\mu \alpha \beta}   \propto 3! ~ \epsilon_{\alpha \beta \mu \rho} \int^1_0 dx ~\{ -2 x (p_{\chi})_\sigma \int \frac{d^4 l}{(2 \pi)^4} \frac{l^\rho l^\sigma}{(l^2-\Delta)^4} -x p_\chi^\rho \cdot \nonumber \\
&  \int  \frac{d^4 l}{(2 \pi)^4}  \frac{l^2}{(l^2-\Delta)^4} -x p_\chi^\rho (x^2 p^2_\chi -m^2) \int  \frac{d^4 l}{(2 \pi)^4}  \frac{1}{(l^2-\Delta)^4} ~.~
\end{align}
Finally, after integrating  out the $l^\mu$, we will get a simple $m_{\tilde{q}}$ and $p_{\chi}$ dependence of $\Gamma^{a b}_{\mu \alpha \beta}$:
\begin{align}
\Gamma^{\text{loop}}(m_{\tilde{q}}, p_{\chi}) = \int^1_0 dx ( \frac{3 x p_\chi^\rho}{\Delta} -  \frac{x p_\chi^\rho (x^2 p_\chi^2 -m^2)}{\Delta^2}) \label{amploop2}
\end{align}

On the other hand, the $m_{\tilde{q}}$ dependence of the tree level effective coupling can be factored out as
\begin{align}
\Gamma^{\text{tree}}  \propto \frac{1}{(m^2 - m^2_\chi)^2- m^2_{\tilde{q}}}~.~
\end{align}
So we can define the ratio
\begin{align}
\text{Ratio} &\propto \Gamma^{\text{loop}} / \Gamma^{\text{tree}}  \nonumber \\
 &= ((m^2 - m^2_\chi)^2- m^2_{\tilde{q}}) \cdot \Gamma^{\text{loop}}(m_{\tilde{q}}, p_{\chi}) ~.~
\end{align}

Taking the non-relativistic limit for the DM momentum, i.e. $p_\chi = (m_\chi,0,0,0)$, $m=m_b$ and $m_{\tilde{q}} = m_{\Delta} +m_\chi$, we solve the ratio numerically. The results are shown in the upper panel of Fig.~\ref{fig:prec}. From the figure we find that the ratio tends to a constant in the heavy squark region for given $m_\chi$, which means the tree level description is accurate. 
While in the region of small mass splitting, the tree level results deviate from the full loop calculation considerably by a amount.  

The range of $m_{\Delta}$  that permits the tree level approximation in required precision $P$ can be solved by using the inequality
\begin{align}
\text{Ratio} (m_\Delta) / \text{Ratio} (m_{\Delta} =500) > (1-P)  \label{rat}
\end{align}
at each given DM mass. 
In the lower panel of Fig.~\ref{fig:prec}, we show the $m_{\Delta}$ region in which the tree level approximation matches the loop level result within 20\% and 50\% precision, respectively. For example, when $m_{\chi} \sim 10$ GeV, $m_{\Delta} \gtrsim 20$ GeV is sufficient to guarantee that  the tree level approximation is accurate within 20\%.

%%%%%%%%%%%%%%%%%%%%%%%%%%%%%%%%%%%%%%%%%%%%%%%%%%%

%\bibliography{SDDM}
%\bibliographystyle{utphys}

\end{document}